\documentclass[onecolumn,prb]{revtex4}
\usepackage{graphicx}
\usepackage{dcolumn}
\usepackage{bm}
\usepackage{amsmath}

\begin{document}

\preprint{}
\title{Charge transport in two dimensional electron  gas/superconductor junctions with Rashba spin-orbit coupling}
\author{T. Yokoyama, Y. Tanaka and J. Inoue }
\affiliation{Department of Applied Physics, Nagoya University, Nagoya, 464-8603, Japan%
\\
and CREST, Japan Science and Technology Corporation (JST) Nagoya, 464-8603,
Japan}
\date{\today}

\begin{abstract}
 We have studied the tunneling conductance in two dimensional electron  gas / insulator / superconductor junctions in the presence of Rashba spin-orbit coupling (RSOC).  It is  found 
that for low insulating barrier the tunneling conductance is suppressed by the RSOC while for high insulating barrier it is almost independent of the RSOC. We also find the reentrant behavior of the conductance at zero voltage as a function of RSOC for intermediate insulating barrier strength. The results are essentially different from those predicted in ferromagnet / superconductor junctions. 
The present derivation of the conductance  is applicable to arbitrary velocity operator with off-diagonal components. 
\end{abstract}

\maketitle



%

%



\section{Introduction}
In normal metal / supercunductor (N/S) junctions Andreev reflection (AR)\cite{Andreev} is one of the most important process for low energy transport. The AR is a process that  an electron with up spin injected from N at the energy below the energy gap $\Delta$ is converted into a reflected hole with up spin. To describe the charge transport in N/S junctions Blonder, Tinkham and Klapwijk (BTK) proposed the formula for the  calculation of the tunneling conductance\cite{BTK}. A gap like structure and the douling of tunneling conductance appear in the voltage dependence due to the AR. This method has been extended to the ferromagnet / superconductor (F/S) junctions and used to estimate the spin polarization of the F layer experimentally\cite{Tedrow,Upadhyay,Soulen}. In F/S junctions, AR is suppressed because the retro-reflectivity is broken by  the exchange field in the F layer\cite{de Jong}. As a result, the conductance of the junctions is suppressed\cite{FS}. Spin dependent transport in F/S junctions is an important subject in the field of spintronics which aims to fabricate  novel devices manipulating electron's spin. 

Spintronics has recently received much attention because of its potential impact on electric devices and quantum computing\cite{Zutic}. Among recent works, many efforts have been devoted to study the effect of spin-orbit coupling on transport properties of two dimensional electron gas (2DEG)\cite{Hirsch,Governale,Streda,Mishchenko,Schliemann,Sinova}. The pioneering work by Datta and Das suggested the way to control the precession of the spins of electrons by the Rashba spin-orbit coupling (RSOC)\cite{Rashba} in F/2DEG/F junctions\cite{Datta}. This spin-orbit coupling depends on the applied electric field and can be tuned by a gate voltage. On the other hand spin dependent transport based only on spin-orbit coupling without ferromagnet, e.g., spin Hall effect is also a hot topic\cite{Edelstein,Inoue,Watson}. 

\begin{figure}[htb]
\begin{center}
\scalebox{0.4}{
\includegraphics[width=30.0cm,clip]{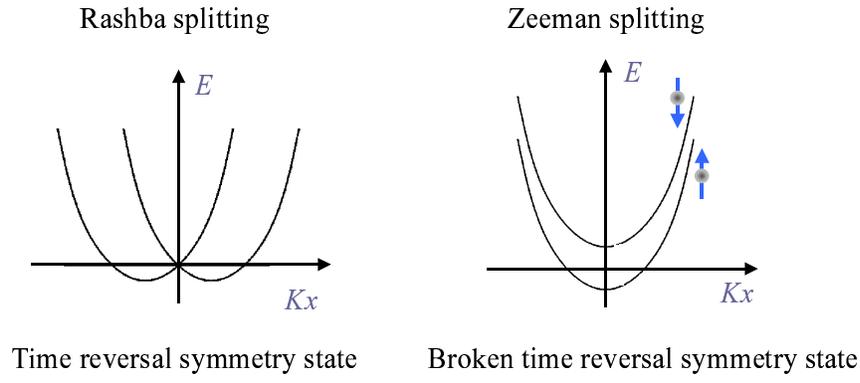}}
\end{center}
\caption{(color online) Schematic illustration of Rashba and Zeeman splitting.} \label{f1}
\end{figure}

As in the case of exchange field in F/S junctions, RSOC may affect the tunneling conductance in 2DEG/S junctions because RSOC mixes spin-up and spin-down states.  The RSOC induces an energy splitting which lifts the spin degeneracy, but the  energy splitting doesn't break the time reversal symmetry unlike an exchange splitting in ferromagnet (see Fig. \ref{f1}). Therefore transport properties in 2DEG/S junctions may be qualitatively different from  those in F/S junctions.  However, in 2DEG/S junctions  the effect of RSOC on transport phenomena is not studied well. It is desirable to make a formalism incorporating the effect of the RSOC in these junctions. For this purpose a BTK-like formula may be accessible. However the derivation of  the conductance by BTK  cannot be directly extended to that in 2DEG/S junctions because velocity operator has off-diagonal components by RSOC. 

 In this paper  
we present a general method to  derive a conductance in superconducting junctions which is applicable to arbitrary velocity operator with off-diagonal components. Applying it, we calculate the tunneling conductance in 2DEG/S junctions, compare it with that in F/S junctions and clarify how  RSOC affects the AR and normal reflection probabilities. 
The obtained results can be useful for the design of mesoscopic 2DEG/S junctions and for a better understanding of related experiments.

The organization of this paper is as follows. In section II, we will provide
the detailed derivation of the expression for the 
conductance. In section III, the results of calculations are presented for
various types of junctions.  In section IV, the summary of the obtained results is given. In the present paper we confine ourselves to zero temperature. 


\section{Formulation}

We consider a ballistic 2DEG / S junctions where
  the 2DEG/S interface is 
located at $x=0$ (along the $y$-axis), and has an infinitely
narrow insulating barrier described by the delta function $U(x)=U\delta
(x)$. 

\begin{figure}[htb]
\begin{center}
\scalebox{0.4}{
\includegraphics[width=28.0cm,clip]{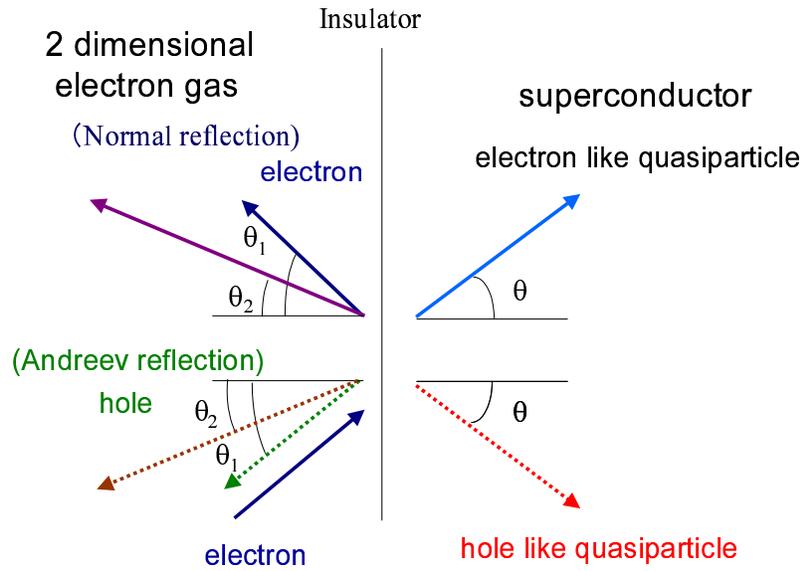}}
\end{center}
\caption{(color online) Schematic illustration of scattering processes.}\label{f2}\end{figure}

The effective Hamiltonian with RSOC is given by
\begin{equation}
H = \left( {\begin{array}{*{20}c}
   {\xi _k } & {i\lambda k_ -  \theta \left( { - x} \right)} & 0 & {\Delta \theta \left( x \right)}  \\
   { - i\lambda k_ +  \theta \left( { - x} \right)} & {\xi _k } & { - \Delta \theta \left( x \right)} & 0  \\
   0 & { - \Delta \theta \left( x \right)} & { - \xi _k } & { - i\lambda k_ +  \theta \left( { - x} \right)}  \\
   {\Delta \theta \left( x \right)} & 0 & {i\lambda k_ -  \theta \left( { - x} \right)} & { - \xi _k }  \\
\end{array}} \right)
\end{equation}
with $k_ \pm   = k_x  \pm ik_y $, the energy gap $\Delta$,  $\xi _k  = \frac{{\hbar ^2 }}{{2m}}\left( {k^2  - k_F^2 } \right)$,  Fermi wave number $k_F$, Rashba coupling constant $\lambda$,  and step function $\theta(x)$.

Velocity operator in the $x$-direction is given by\cite{Molenkamp}
\begin{equation}
v_x  = \frac{{\partial H}}{{\hbar \partial k_x }} = \left( {\begin{array}{*{20}c}
   {\frac{\hbar }{{mi}}\frac{\partial }{{\partial x}}} & {\frac{{i\lambda }}{\hbar }\theta \left( { - x} \right)} & 0 & 0  \\
   { - \frac{{i\lambda }}{\hbar }\theta \left( { - x} \right)} & {\frac{\hbar }{{mi}}\frac{\partial }{{\partial x}}} & 0 & 0  \\
   0 & 0 & { - \frac{\hbar }{{mi}}\frac{\partial }{{\partial x}}} & { - \frac{{i\lambda }}{\hbar }\theta \left( { - x} \right)}  \\
   0 & 0 & {\frac{{i\lambda }}{\hbar }\theta \left( { - x} \right)} & { - \frac{\hbar }{{mi}}\frac{\partial }{{\partial x}}}  \\
\end{array}} \right).
\end{equation}

As shown in Fig. \ref{f2}, the wave function $\psi(x)$  for $x \le 0$ is represented using eigenfunctions of the Hamiltonian: 
\begin{equation}
\begin{array}{l}
 \psi(x \le 0) = e^{ik_y y} \left[ {\frac{1}{{\sqrt 2 }}e^{ik_{1(2)} \cos \theta _{1(2)} x} \left( {\begin{array}{*{20}c}
   {\left(  -  \right)i\frac{{k_{1(2) - } }}{{k_{1(2)} }}}  \\
   1  \\
   0  \\
   0  \\
\end{array}} \right) + \frac{{a_{1(2)} }}{{\sqrt 2 }}e^{ik_1 \cos \theta _1 x} \left( {\begin{array}{*{20}c}
   0  \\
   0  \\
   {i\frac{{k_{1 + } }}{{k_1 }}}  \\
   1  \\
\end{array}} \right) + \frac{{b_{1(2)} }}{{\sqrt 2 }}e^{ik_2 \cos \theta _2 x} \left( {\begin{array}{*{20}c}
   0  \\
   0  \\
   { - i\frac{{k_{2 + } }}{{k_2 }}}  \\
   1  \\
\end{array}} \right)} \right. \\ 
 \left. { + \frac{{c_{1(2)} }}{{\sqrt 2 }}e^{ - ik_1 \cos \theta _1 x} \left( {\begin{array}{*{20}c}
   { - i\frac{{k_{1 + } }}{{k_1 }}}  \\
   1  \\
   0  \\
   0  \\
\end{array}} \right) + \frac{{d_{1(2)} }}{{\sqrt 2 }}e^{ - ik_2 \cos \theta _2 x} \left( {\begin{array}{*{20}c}
   {i\frac{{k_{2 + } }}{{k_2 }}}  \\
   1  \\
   0  \\
   0  \\
\end{array}} \right)} \right] 
 \end{array}
\end{equation}
for an injection wave with wave number $k_{1(2)}$ where 
 $ k_1  =  - \frac{{m\lambda }}{{\hbar ^2 }} + \sqrt {\left( {\frac{{m\lambda }}{{\hbar ^2 }}} \right)^2  + k_F^2 } $,
$ k_2  =   \frac{{m\lambda }}{{\hbar ^2 }} + \sqrt {\left( {\frac{{m\lambda }}{{\hbar ^2 }}} \right)^2  + k_F^2 } $ and $k_{1(2) \pm }  = k_{1(2)} e^{ \pm i\theta _{1(2)} } $. $a_{1(2)}$ and $ b_{1(2)}$ are AR coefficients. $c_{1(2)}$ and $d_{1(2)}$ are normal reflection coefficients.
 $\theta _{1(2)}$ is an angle of the wave with wave number $k_{1(2)}$ with respect to the interface normal.

Similarly for $x \ge 0$ $\psi(x)$ is given by
\begin{equation}
\psi(x \ge 0) = e^{ik_y y} \left[ {e_{1(2)} e^{ik_F \cos \theta x} \left( {\begin{array}{*{20}c}
   u  \\
   0  \\
   0  \\
   v  \\
\end{array}} \right) + f_{1(2)} e^{ik_F \cos \theta x} \left( {\begin{array}{*{20}c}
   0  \\
   u  \\
   { - v}  \\
   0  \\
\end{array}} \right) + g_{1(2)} e^{ - ik_F \cos \theta x} \left( {\begin{array}{*{20}c}
   v  \\
   0  \\
   0  \\
   u  \\
\end{array}} \right) + h_{1(2)} e^{ - ik_F \cos \theta x} \left( {\begin{array}{*{20}c}
   0  \\
   { - v}  \\
   u  \\
   0  \\
\end{array}} \right)} \right]
\end{equation}
with 
\begin{equation}
u = \sqrt {\frac{1}{2}\left( {1 + \frac{{\sqrt {E^2  - \Delta ^2 } }}{E}} \right)}, \quad \quad v = \sqrt {\frac{1}{2}\left( {1 - \frac{{\sqrt {E^2  - \Delta ^2 } }}{E}} \right)} 
\end{equation}
 where $E$ is quasiparticle energy and $\theta$ is an angle of the wave with wave number $k_{F}$ with respect to the interface normal. $e_{1(2)},  f_{1(2)},  g_{1(2)}$ and $h_{1(2)}$ are transmission coefficients. Note that since the translational symmetry holds for the $y$-direction, the momenta parallel to the interface are conserved: $k_y=k_F \sin \theta  = k_1 \sin \theta _1  = k_2 \sin \theta _2 $.

 The wave function follows the boundary conditions\cite{Molenkamp}:
\begin{equation}
\begin{array}{l}
 \left. {\psi \left( x \right)} \right|_{x =  + 0}  = \left. {\psi \left( x \right)} \right|_{x =  - 0}  \\ 
 \left. {v_x \psi \left( x \right)} \right|_{x =  + 0}  - \left. {v_x \psi \left( x \right)} \right|_{x =  - 0}  = \frac{\hbar }{{mi}}\frac{{2mU}}{{\hbar ^2 }}\tau _3 \psi \left( 0 \right) \\ 
 \tau _3  = \left( {\begin{array}{*{20}c}
   1 & 0 & 0 & 0  \\
   0 & 1 & 0 & 0  \\
   0 & 0 & { - 1} & 0  \\
   0 & 0 & 0 & { - 1} 
\end{array}} \right).
 \end{array}
\end{equation}

Now we will derive a formula for the tunneling conductance. Before giving detailed calculation, we present an essential idea for the derivation of the conductance: First we calculate  expectations of current for the complete sets of the eigenfunctions. Next we sum the expectations multiplied by corresponding distribution functions. Then we can get the net current of the junctions. Detailed derivation is given in the Appendix. 

 Finally we obtain the dimensionless conductance  represented in the form:
\begin{equation*}
\begin{array}{l}
 \sigma _s  = N_1 \int_{ - \theta _C }^{\theta _C } {\frac{1}{2}\left[ {\left( {1 + \frac{{k_2 }}{{k_1 }}} \right) + \left| {a_1 } \right|^2 \left( {1 + \frac{{k_2 }}{{k_1 }}} \right) + \left| {b_1 } \right|^2 \left( {1 + \frac{{k_1 }}{{k_2 }}} \right)\lambda _{21}  - \left| {c_1 } \right|^2 \left( {1 + \frac{{k_2 }}{{k_1 }}} \right) - \left| {d_1 } \right|^2 \left( {1 + \frac{{k_1 }}{{k_2 }}} \right)\lambda _{21} } \right]} \cos \theta d\theta  \\ 
  + N_2 \int_{ - \frac{\pi }{2}}^{\frac{\pi }{2}} {{\mathop{\rm Re}\nolimits} \frac{1}{2}\left[ {\left( {1 + \frac{{k_1 }}{{k_2 }}} \right) + \left| {a_2 } \right|^2 \left( {1 + \frac{{k_2 }}{{k_1 }}} \right)\lambda _{12}  + \left| {b_2 } \right|^2 \left( {1 + \frac{{k_1 }}{{k_2 }}} \right) - \left| {c_2 } \right|^2 \left( {1 + \frac{{k_2 }}{{k_1 }}} \right)\lambda _{12}  - \left| {d_2 } \right|^2 \left( {1 + \frac{{k_1 }}{{k_2 }}} \right)} \right]} \cos \theta d\theta  \\ 
  = \int_{ - \theta _C }^{\theta _C } {\left[ {1 + \left| {a_1 } \right|^2  + \left| {b_1 } \right|^2 \frac{{k_1 }}{{k_2 }}\lambda _{21}  - \left| {c_1 } \right|^2  - \left| {d_1 } \right|^2 \frac{{k_1 }}{{k_2 }}\lambda _{21} } \right]} \cos \theta d\theta  
 \\ + \int_{ - \frac{\pi }{2}}^{\frac{\pi }{2}} {{\mathop{\rm Re}\nolimits} \left[ {1 + \left| {a_2 } \right|^2 \frac{{k_2 }}{{k_1 }}\lambda _{12}  + \left| {b_2 } \right|^2  - \left| {c_2 } \right|^2 \frac{{k_2 }}{{k_1 }}\lambda _{12}  - \left| {d_2 } \right|^2 } \right]} \cos \theta d\theta  \\ 
 \end{array}
\end{equation*}
\begin{equation}
 \equiv  \left( {1 + A_1  + B_1  + C_1  + D_1 } \right)\int_{ - \theta _C }^{\theta _C } {\cos \theta d\theta } + 2\left( {1 + A_2  + B_2  + C_2  + D_2 } \right) 
\end{equation}
where
\begin{equation}
N_1  =\frac{1}{{1 + \frac{{m\lambda }}{{\hbar ^2 k_1 }}}} \quad \quad N_2  = \frac{1}{{1 - \frac{{m\lambda }}{{\hbar ^2 k_2 }}}}.
\end{equation}
$N_1$ and $N_2$ are  density of states normalized by those with $\lambda=0$ for wave number $k_1$ and $k_2$ respectively. $\lambda _{12}$ and $\lambda _{21}$ are defined in the Appendix. 
The critical angle $\theta _C$ is defined as $\cos \theta _C  = \sqrt {\frac{{2m\lambda }}{{\hbar ^2 k_1 }}} $. 

$\sigma _N$ is given by the conductance for normal states, i.e., $\sigma _S$ for $\Delta=0$. We define normalized conductance as $\sigma _T =\sigma _S /\sigma _N$ and parameters as
$\beta  = \frac{{2m\lambda }}{{\hbar ^2 k_F }}$ and 
$Z = \frac{{2mU}}{{\hbar ^2 k_F }}$. For example, in InGaAs heterostructures, $\beta$ is estimated as $\beta \sim 0.2$.\cite{Grundler,Sato}
 Here we choose the same effective mass in 2DEG and S. In most cases the effective mass in 2DEG is much smaller than that in S. However we can show that this effect is equivalent to that by the increase of  Z. Thus we neglect the difference of the effective masses in the present paper. 

\section{Results}
\begin{figure}[htb]
\begin{center}
\scalebox{0.4}{
\includegraphics[width=18.0cm,clip]{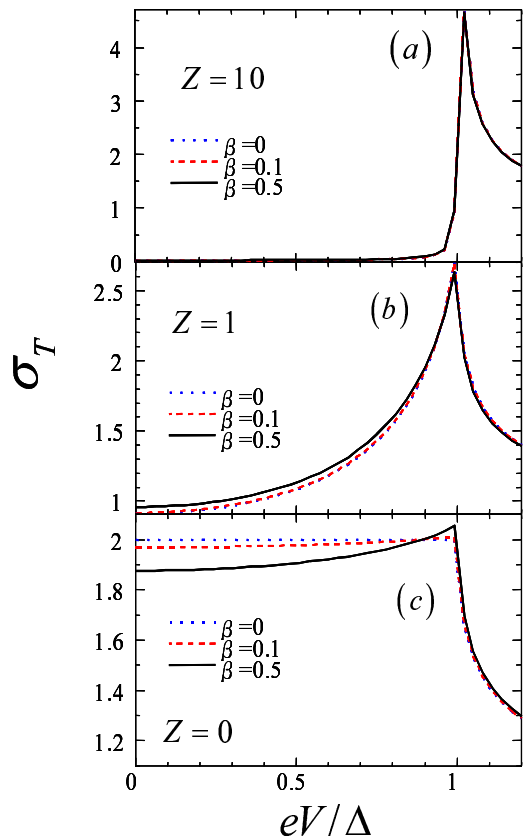}}
\end{center}
\caption{(color online) Normalized tunneling conductance with $Z=10$ in (a), $Z=1$ in (b), and  $Z=0$ in (c).} \label{f3}
\end{figure}

First we study the normalized tunneling conduntace $\sigma _T$ as a function of bias voltage $V$ in Fig. \ref{f3}.
For $Z=10$ where the AR probability is low, $\sigma _T$ is almost zero within the enregy gap and independent of $\beta$. In contrast, for $Z=1$, $\sigma _T$ is slightly enhanced with the increase of $\beta$ around zero voltage. For $Z=0$ where the AR probability is very high, $\sigma _T$ becomes two for  $\beta=0$ within the energy gap. It is  reduced by the increase  $\beta$ within the energy gap.

\begin{figure}[htb]
\begin{center}
\scalebox{0.4}{
\includegraphics[width=28.0cm,clip]{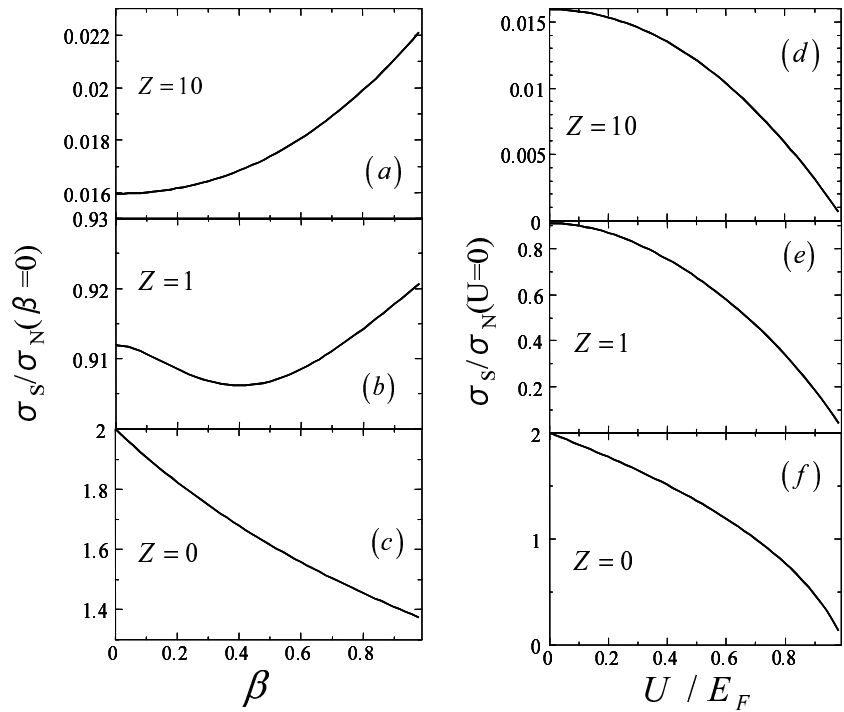}}
\end{center}
\caption{ Tunneling condutance for superconducting states  at zero voltage as a function of  RSOC in 2DEG/S junctions  (left panels) and  the exchange field in F/S junctions (right panels) with $Z=10$ in (a) and (d), $Z=1$ in (b) and (e), and  $Z=0$ in (c) and (f). Here  $\beta  = \frac{{2m\lambda }}{{\hbar ^2 k_F }}$.} \label{f8}
\end{figure}

Next we study the difference between the effect of the Rashba splitting and that of  Zeeman splitting. We have calculated conductance in F/S junctions following Ref. \cite{FS}. 
We plot the tunneling condutance for superconducting states $\sigma _S$ at zero voltage for 2DEG/S junctions in (a)-(c) and for F/S junctions in (d)-(f) of Fig. \ref{f8} with $Z=10$ in (a) and (d), $Z=1$ in (b) and (e), and  $Z=0$ in (c) and (f). 
In (a)-(c) we show the dependence of $\sigma _S$, normalized by $\sigma _N$ for $\beta=0$,  on $\beta$ for various $Z$. For $Z=10$ it has an exponential dependence on $\beta$ but its magnitude is very small while it has a reentrant behavior as a function of $\beta$ for $Z=1$. For $Z=0$ it decreases linearly as a function of $\beta$. 
On the other hand, in F/S junctions, the dependence of $\sigma _S$ on $U$, normalized by Fermi energy $E_F$, is qualitatively different. We plot $\sigma _S$  normalized by $\sigma _N$ at $U=0$. 
The exchange field suppresses  $\sigma _S$ independently of $Z$ as shown in (d)-(f). This is because the AR probability is reduced by the  exchange field. 
Therefore the effect of the Rashba splitting on conductance is essentially different from that of  Zeeman splitting on conductance. This can be explained as follows. The Zeeman splitting gives unbalance of populations of up and down spin electrons. Thus it suppresses the AR where pairs of spin-up and spin-down electrons are transmitted to S. On the other hand, the Rashba splitting never causes such an unbalance. Thus it cannot suppresses the AR, which results in various $\beta$ dependence of the conductance. 

\begin{figure}[htb]
\begin{center}
\scalebox{0.4}{
\includegraphics[width=27.0cm,clip]{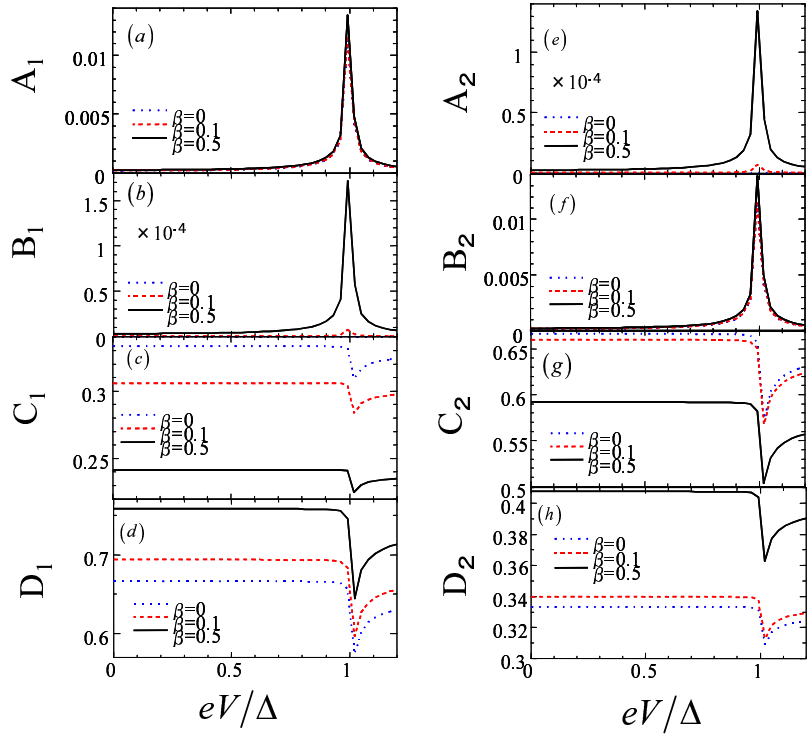}}
\end{center}
\caption{(color online) The angular averaged  Andreev and normal reflection probabilities for $Z=10$. $A_1$, $A_2$, $B_1$ and $B_2$ are AR probabilities.  $C_1$, $C_2$, $D_1$ and $D_2$ are normal reflection probabilities. } \label{f6}
\end{figure}

\begin{figure}[htb]
\begin{center}
\scalebox{0.4}{
\includegraphics[width=26.0cm,clip]{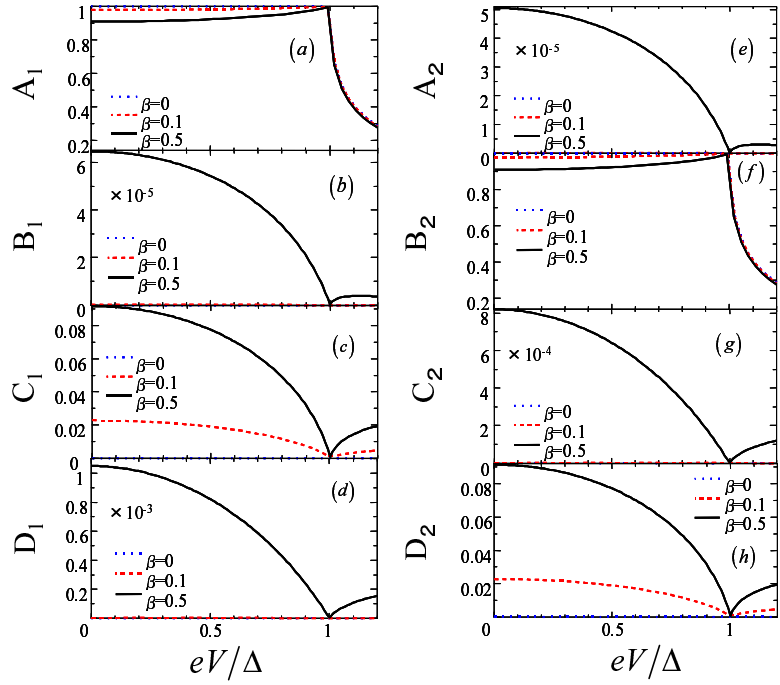}}
\end{center}
\caption{(color online) The angular averaged  Andreev and normal reflection probabilities for $Z=0$.  $A_1$, $A_2$, $B_1$ and $B_2$ are AR probabilities.  $C_1$, $C_2$, $D_1$ and $D_2$ are normal reflection probabilities. } \label{f7}
\end{figure}

In order to explain the line shapes of the conductances, we will check the angular averaged normal reflection and AR probabilities as a function of voltage. 
For large $Z$, AR probabilities are small and  normal reflection probabilities reflect the dependence of  the densities of states on $\beta$: $N_1$ is a decreasing function of  $\beta$, while $N_2$ is an increasing function of  $\beta$. Therefore  $C_1$ and $C_2$ are reduced and  $D_1$ and $D_2$ are enhanced with the increase of  $\beta$. 
Figure \ref{f6} shows the probabilities for $Z=10$. AR probabilities ($A_1$, $A_2$, $B_1$ and $B_2$) are slightly enhanced around $eV=\Delta$ and have similar structures  with the increase of $\beta$ while normal reflection probabilities $D_1$ and $D_2$ increase with the increase of $\beta$.  On the other hand normal reflection probabilities  $C_1$ and $C_2$ are reduced with the increase of $\beta$. In other words, reflected waves with wave number $k_1 (k_2)$ are suppressed (enhanced) by RSOC. The enhancement and the suppression compete with  each other. Thus the conductance is almost independent of RSOC. 
For small $Z$,  normal reflection probabilities are small. AR probabilities $A_1$ and $B_2$ are reduced as increasing $\beta$. This stems from the mismatch of Fermi surfaces between 2DEG and S by the increase of $\beta$. In fact, for $Z=0$ (see Fig. \ref{f7}) normal reflection probabilities $C_2$  and $D_1$, and AR probabilities $B_1$ and $B_2$  are slightly  enhanced with the increase of $\beta$. Normal reflection probabilities $C_1$ and  $D_2$ increase by the increase of $\beta$. On the other hand AR probabilities $A_1$ and $B_2$  within the energy gap are reduced with the increase of $\beta$. This means that only eigenfunctions with the same wave number as the injection wave are affected by  RSOC.
 From Fig. \ref{f7}, we can understand the suppression of the tunneling conductance  by  RSOC (see Eq. (7)). 

\clearpage
\section{Conclusions}

In the present paper we have studied the tunneling conductance in two dimensional electron  gas / insulator / superconductor junctions with RSOC.  We have extended the  BTK formula and calculated the tunneling conductance. It is found 
that for low insulating barrier the tunneling conductance is suppressed by the RSOC while for high insulating barrier the tunneling conductance is almost independent of it. We also found a reentrant behavior of the conductance at zero voltage as a function of RSOC for intermediate insulating barrier strength. This phenomena are essentially different from those found in F/S junctions where the tunneling conductance is suppressed by exchange field, being independent of the barrier strength. 
The present derivation of the conductance is applicable to arbitrary velocity operator with off-diagonal components. 

The results give  the possibility to control the AR probability by a gate voltage. We believe that the obtained results are useful for the design of   mesoscopic 2DEG/S junctions and for a better understanding of related experiments.

In this paper we focus on ballistic 2DEG/S junctions. In diffusive 2DEG/S junctions, proximity effect plays an important role. The RSOC breaks the inversion symmetry and hence mixes the parity. As a result, "triplet" pairing may be induced in the 2DEG region\cite{Edelstein2} as predicted in diffusive F/S junctions\cite{Bergeret}. The study in this direction  is now in progress. 


%
The authors appreciate useful and fruitful discussions with  A. Golubov.
This work was supported by
NAREGI Nanoscience Project, the Ministry of Education, Culture,
Sports, Science and Technology, Japan, the Core Research for Evolutional
Science and Technology (CREST) of the Japan Science and Technology
Corporation (JST) and a Grant-in-Aid for the 21st Century COE "Frontiers of
Computational Science" . The computational aspect of this work has been
performed at the Research Center for Computational Science, Okazaki National
Research Institutes and the facilities of the Supercomputer Center,
Institute for Solid State Physics, University of Tokyo and the Computer Center.

\section*{Appendix}
Here we give a detailed derivation of the conductance which is applicable to arbitrary velocity operator with off-diagonal components. 
For an electron injection and a hole injection from 2DEG (represented by $\psi _e$ and $\psi _h$, respectively),  the resulting currents $j_e$ and $j_h$ in the 2DEG region are given by 
\begin{equation}
 j_e  = {\mathop{\rm Re}\nolimits} (\psi_e^\dag  v_x \tau _3 \psi _e ) \propto \left( {1 + A^{he}  + B^{he}  - C^{ee}  - D^{ee} } \right)
\end{equation}
\begin{equation}
 j_h  = {\mathop{\rm Re}\nolimits} (\psi _h ^\dag  v_x \tau _3 \psi _h ) \propto \left( {1 + A^{eh}  + B^{eh}  - C^{hh}  - D^{hh} } \right).
\end{equation}
Similary, for an electron  and a hole injection  from S (represented by $\psi{\prime} _e$ and $\psi{\prime} _h$, respectively), the corresponding currents $j_e ^\prime$ and  $j_h ^\prime$  in the 2DEG region reads 
\begin{equation}
 j_e ^\prime   = {\mathop{\rm Re}\nolimits} (\psi_{e }^{\prime \dag}  v_x \tau _3 \psi '_e ) \propto \left( {F^{ee}  + G^{ee}  - H^{he}  - J^{he} }\right) 
\end{equation}
\begin{equation}
 j_h ^\prime   = {\mathop{\rm Re}\nolimits} (\psi _{h }^{\prime \dag}  v_x \tau _3 \psi '_h ) \propto \left( {F^{hh}  + G^{hh}  - H^{eh}  - J^{eh} } \right).
\end{equation}
Here $A^{he}$ and $B^{he}$, and, $C^{ee}$ and $D^{ee}$ denote AR and normal reflection probabilities  with electron injection  respectively. 
$F^{ee}$ and   $G^{ee}$  are normal  transmission probabilities with electron injection. 
$H^{he}$ and $J^{he}$ are  transmission probabilities with the injection of an electron converted into a hole at the interface. 
Other notations are defined in a similar way.  
Note that there are four independent eigenfunctions: two kinds of electron-like quasiparticles and two kinds of hole-like quasiparticles. 
The total current reads 
\begin{equation*}
 I = \int_{ - \infty }^\infty  {\left( {j_e f(E - eV) - j_h f(E + eV) + j_e ^\prime  f(E) - j_h ^\prime  f(E)} \right)} dE 
\end{equation*}
\begin{equation}
 \propto \int_{ - \infty }^\infty  {\left( {1 + A_1^{he}  + B_1^{he}  - C_1^{ee}  - D_1^{ee} } \right)\left( {f(E - eV) - f(E + eV)} \right)} dE.
\end{equation}
Then the differential conductance at zero temperature has the form: 
\begin{equation}
 \frac{{dI}}{{dV}} \propto \left( {1 + A^{he}  + B^{he}  - C^{ee}  - D^{ee} } \right) 
\end{equation}
with Fermi distribution function $f(E)$ and bias voltage $V$. 
Here we assume the particle-hole symmetry which results in the relations $X^{ee}=X^{hh}$ and $ X^{he}=X^{eh} (X=A,B,C,D,F,G,H,J)$. The original BTK method\cite{BTK} cannot treat velocity operator with off-diagonal components. However, 
the derivation given here is applicable to arbitrary velocity operator with off-diagonal components. 

Let us apply the above procedure to our model. For an injection wave with wave number $k_{1}$, the current reads 
\begin{equation}
 j_e^1  = \frac{{\hbar ^2 }}{{2m}}\left( {1 + \frac{{k_2 }}{{k_1 }}} \right)k_1 \cos \theta _1 \left( {1 + \left| {a_1 } \right|^2  + \left| {b_1 } \right|^2 \frac{{k_1 }}{{k_2 }}\lambda _{21}  - \left| {c_1 } \right|^2  - \left| {d_1 } \right|^2 \frac{{k_1 }}{{k_2 }}\lambda _{21} } \right). 
\end{equation}
For an injection wave with wave number $k_{2}$, the current is 
\begin{equation}
 j_e^2  = \frac{{\hbar ^2 }}{{2m}}\left( {1 + \frac{{k_1 }}{{k_2 }}} \right)k_2 \cos \theta _2 \left( {1 + \left| {a_2 } \right|^2 \frac{{k_2 }}{{k_1 }}\lambda _{12}  + \left| {b_2 } \right|^2  - \left| {c_2 } \right|^2 \frac{{k_2 }}{{k_1 }}\lambda _{12}  - \left| {d_2 } \right|^2 } \right). 
\end{equation}
Here we define $\lambda _{12}$ and $\lambda _{21}$ as 
\begin{equation}
\lambda _{12}  = \frac{{k_1 \cos \theta _1 }}{{k_2 \cos \theta _2 }} \quad \quad \lambda _{21}  = \frac{{k_2 \cos \theta _2 }}{{k_1 \cos \theta _1 }}. 
\end{equation}

%


\end{document}